\documentstyle[12pt,psfig]{article}

\newcommand{\beq}[1]{\begin{equation} \label{#1}}
\newcommand{\eeq}{\end{equation}}
\newcommand{\beqa}[1]{\begin{eqnarray} \label{#1}}
\newcommand{\eeqa}{\end{eqnarray}}
\newcommand{\beqn}{\begin{eqnarray*}}
\newcommand{\eeqn}{\end{eqnarray*}}

\newcommand{\Dt}{\frac{d}{d{t}}}

\newcommand{\rf}[1]{(\ref{#1})}

\newcommand{\veps}{\varepsilon}

\newcommand{\ga}{\alpha}
\newcommand{\gb}{\beta}
\newcommand{\gl}{\lambda}

\newcommand{\gvk}{\ae}

\newcommand{\Gg}{\gamma}

\newcommand{\lln}{{\rm ln}}
\newcommand{\const}{{\rm const}}

\title{The cosmological model with an analytic exit from inflation.}
\author{Chervon, S.V.,$^{(1)}$ Zhuravlev, V.M.$^{(2)}$}

\begin{document}
\maketitle

\begin{center} (1) Department of Theoretical and Mathematical Physics, Ulyanovsk State
University,\\
42, Leo Tolstoy, Ulyanovsk, 432700, Russia\\
{\rm Email: chervon@sv.uven.ru}\\
(2) Department of Theoretical and Mathematical Physics, Ulyanovsk State
University,\\
42, Leo Tolstoy, Ulyanovsk, 432700, Russia\\
{\rm Email: zhuravl@sv.uven.ru}
\end{center}

\begin{abstract}
A cosmological model of homogeneous and isotropic spatially
flat Universe with gravitating self-interacting scalar field is considered.
The exact solution, admitting an analytical exit from
inflationary stage into a radiation era and a matter dominated epoch, is
obtained by virtue of ``fine turning of the potential'' method.
We found that an inflationary stage is supported by decay of higgs bosons
in the framework of the solution obtained. Freidmann's regim is associated
with adiabatical expantion of the Universe, filled by the matter with special
equation of state.

Thus we presented the exact solution which solve the
problem of transition from an inflationary to a radiation eras or long
standing `exit' problem.
\end{abstract}

\bigskip

PACS-94: 04.20.-q \ 98.80.-Dr
\bigskip

\section{Introduction}
In this contribution we investigate general properties of the evolution of a
homogeneous isotropic universe on the basis of an analysis of the system of
Einstein equations for a self-inter\-acting scalar field.  One of the
important requirements imposed on early universe theories by present-day
observational data is the existence of an inflationary phase in the
evolution of the universe with the scale factor increasing at an
exponential rate.\cite{linde90}
Thus the morden general consept of universe's evolution include the
evolution from the Planck's epoch through an inflationary stage which should
be ended by exit to radiation era and then to matter dominated epoch to the
present-day universe.
In a literature we can find a lot of theories which can describe all of
mentioned epochs, but separately, that is the the physical content of the
model is changed from one regim to enother. Therefore
it is important to construct the model of one physical content
describing the global evolution of the Universe starting
from the preinflationary epoch and ending with later epochs in the
evolution of the universe. The model of this kind should have a solution of
`exit problem' from inflation stage to radiation and matter dominated
stage, mentioned in the original work of A.Guth \cite{guth81} as a central
problem of inflationary scenario.
In the present work we present
the exact solution, admitting an analytical exit from
inflationary stage into a radiation era and a matter dominated epoch. The
result is obtained by virtue of ``fine turning of the potential'' method
and by the analysis of some correspondence between exact solutions and
approximate ones in slow roll regim.
We found that an inflationary stage is supported by decay of higgs bosons
in the framework of the solution obtained. Freidmann's regim is associated
with adiabatical expantion of the Universe, filled by the matter with special
equation of state.

\section{Survey of exact solutions in inflation}
The inflationary phase of the (isotropic and homogeneous) universe is
customarily understood to mean the period $t_{i} < t < t_{f}$ of its early
evolution, beginning at the time $t_{i}$ and ending at the time $t_{f}$,
when the scale factor $R(t)$ increased at an accelerated rate, i.e.,
\begin{eqnarray*}
\ddot R=R(H^2+\dot H)>0
\end{eqnarray*}
(see \cite{3}).  In this setting the term subinflation is
used when $\dot{H} < 0$, standard or exponential inflation when $\dot{H} =
0$, and superinflation in the case $\dot{H} > 0$.

We take a special look at the case in which the law governing the variation
of the scale factor $R(t)$ has one of the forms
\begin{eqnarray}
\label{e1} &&R(t) = R_0t^m, \quad m>1, R_0>0,\\
 \label{e2} &&R(t) = R_0e^{H_0t}, \quad H_0>0, \\
 \label{e3} &&R(t) = R_0(t_*-t)^n, \quad n<0, \quad t_*>t_i,
\end{eqnarray}
which establish the general inflation condition $\ddot{R} > 0$.  The scale
factor (\ref{e1}) corresponds to subinflation and is called power-law
inflation.  The case of superinflation (\ref{e3}) is unattainable within
the scope of self-inter\-acting scalar field theory.

The existence of an inflationary period of expansion of the universe is
necessary for solving the horizon (homogeneity), flatness, and relic
monopole problems, which have been discussed in detail, for example, in the
original work of Guth \cite{guth81} (see also Linde's book \cite{linde90}).  The specific
mechanism driving inflation in the theory with a self-inter\-acting scalar
field (inflaton field) is usually identified with the form of the
functional dependence of the effective self-inter\-action potential of the
scalar field on the field itself: $V = V(\phi )$.  An important factor here
in substantiating the existence of such a mechanism is the slow-roll
regime, i.e., a regime in which the variation of the field $\phi $ [and the
potential $V(\phi )$] was relatively slow, so that the ``kinetic'' energy
of the scalar field can be disregarded: $\dot{\phi }^{2} \ll V(\phi )$
(\cite{5}).

In the standard model (\ref{e2}) the self-inter\-action potential is
specified as a quadratic function of~$\phi $:
\begin{eqnarray*}
V(\phi)=\mu^2\phi^2+b.
\end{eqnarray*}
For this case the field varies linearly with time: $\phi \propto t$,
whereas the scale factor increases exponentially: $R(t) \propto e^{Ht}$
(\cite{linde90} and \cite{6}).

A new approach to the construction of exact solutions in inflationary
cosmology (without any reliance on the slow-roll regime) has been developed
in \cite{chezhu96iv,7}, and \cite{ch97gc}.  The method can be broadly
summarized as follows.  We specify the regime of evolution of the scale
factor $R = R(t)$ and determine the evolution of the self-inter\-action
potential $V = V(\phi )$ in such a way as to ensure the specified $R =
R(t)$ regime.  We then determine the rate of change of the scalar field
$\dot{\phi }(t)$.  Next, integrating over {\it t}, we find the law of
evolution of the scalar field $\phi = \phi (t)$, which will also be matched
with the regime chosen for the scale factor.  In the final analysis we
obtain a parametric dependence $V = V(\phi )$, which then represents the
self-inter\-action potential ``fine-tuned'' to the given exact solution.

In previous work, \cite{chezhu96iv} in particular, the self-inter\-action potentials
have
been plotted as a function of the scale factor $R(t)$ corresponding to the
inflationary regimes (\ref{e1}) and (\ref{e2}).  Also in
\cite{chezhu96iv} $V(\phi )$ has been obtained in general parametric
form for certain types of large-{\it t} asymptotic behavior of $R(t)$ such
as to ensure regime (\ref{e1}) or (\ref{e2}).

A somewhat different approach to the construction of exact solutions in
cosmological inflation models has been presented by Barrow.\cite{11}
The crux
of his approach is that the evolution of the scalar field $\phi = \phi (t)$
is specified; the next step is to determine the evolution of the scale
factor $R = R(t)$ and the potential, which depends explicitly on the scalar
field~$\phi $.

We also mention the work of Maartens et al.,\cite{mtr95} who have used a method
similar to fine tuning of the potential, but with a special parameter
introduced in place of time.  This approach might well be practical for the
solution of some problems, but it complicates the process of obtaining and
analyzing solutions.

Here we follow the general scheme of the potential fine-tuning method and
submit two new techniques for constructing classes of exact inflationary
solutions for a homogeneous and isotropic universe.  One technique entails
a variational formulation of the conditions underlying the slow-roll
regime, and the second rests on the possibility of specifying the
self-inter\-action potential as a function of time.  We investigate the
general implications of such an approach and present new exact solutions.
We proposed as well the cosmological model based on exact solutions in which
the quality solution of the global evolution of the universe is found.

The main attention is consentrated on the solution of the exit problem from
inflation, because the trasition from radiation- to matter-dominated epochs
have beed well investigated \cite{zelnov75, weinberg75}.

Let us remember that the exit from inflationary stage should be accompanied
by the set of Freidmann's epoch when the scalar factor has the power law
evolution $R(t)\propto t^s,$ where $~s<1$.
Besides $s=1/2$ -- when radiation-dominated and  $s=2/3$ when matter-dominated
epochs are realized.
Let us mention also that each value of $s$ correponds to certain equation of
state in standard theory of Big Bang \cite{zelnov75}.

For the models of the early univers's evolution, based on the theory of
self-interacting scalar field, the inflationary regim is not specific one
because it can be provided by a wide class of the potentials of
self-interaction \cite{czs97,ZCS98}. Nevertheless the condition of transition
from inflationary regim to freidmann's stages is a specific one and can be
realized under very special restrictions.

We whould like to mention here again the work \cite{mtr95} where the
inflationary
model with an analytical exit to radiation stage and dust matter is proposed.
As a difference from our approuch, the dependence of the Hubbll's `constant'
$H \equiv {\dot R}/{R}$ of $\ln R$ is used in \cite{mtr95};
the e-fold number is chosen as one more dynamical variable as well.

\section{Exact solution and self-inter\-action potential for the slow-roll
regime}\label{s2}

The Einstein equations for a homogeneous isotropic universe, given an
arbitrary form of the self-inter\-action potential $V(\phi )$ of the scalar
field $\phi (t)$, can be represented by two equations in the class of
Friedmann metrics \cite{chezhu96iv}:
\begin{eqnarray}
\label{e4} && V(t)=\frac{1}{\kappa}\left(\Lambda+\frac{\ddot{R}}{R}+2
\frac{\dot{R}^2}{R^2} +\frac{2\epsilon}{R^2}\right),\\
 \label{e5} && \phi(t)=\pm \sqrt{\frac{2}{\kappa}}\int\sqrt{- \frac{d^2{
\rm ln}{R}}{dt^2} +\frac{2\epsilon}{R^2}}dt+\phi_0,
\end{eqnarray}
where $\kappa $ is the gravitational constant, $\Lambda $ is the
cosmological constant, and $\phi _{0}$ is a constant of integration.  This
system of equations provides the basis of the potential fine-tuning method.
As mentioned in the Introduction, once the evolution of the scale factor
has been specified, this method can be used to find the appropriate form of
the self-inter\-action potential to achieve the given regime of evolution
of the universe.

The method has been used previously \cite{chezhu96iv,ch97m} to analyze inflationary regimes and
to demonstrate the existence of a large number of self-inter\-action
potentials admitting such an evolution.  The enormous diversity of
potentials leading to inflation compels us to look for an additional
principle that might be used to isolate the potential that actually
occurred in the early stages of evolution of the universe.  Considering the
special importance of the existence of the slow-roll regime for an
inflationary scenario, we look into a nonstandard formulation of the
slow-roll regime.  We define the slow-roll regime with the aid of the
variational principle of minimum variation of the scalar field $\phi $ with
variation of the scale factor $R(t)$ and, as a consequence, minimum
variation of the values of the potential $V(\phi )$.  On the basis of
Eq.\,\,(\ref{e5}) this condition can be specified by the variational
equation
\begin{eqnarray}
\label{e6} &&\delta\phi=\int\limits_{t_1}^{t_2}\delta\sqrt{- \frac{d^2{
\rm ln}{R}}{dt^2} +\frac{2\epsilon}{R^2}}dt=0,
\end{eqnarray}
where $t_{i} \le t_{1} < t_{2} \le t_{f}$.  This condition
literally means that the evolution of the scale factor must be such that
the difference in the values of the field $\phi $ in any finite time
interval $[t_{1},\, t_{2}]$ will be the smallest among all other possible
evolutions.  The variation of the values of the potential $V(\phi )$ is
also the smallest in this case.

If we introduce the notation
\begin{eqnarray*}
&& F(t)=\left(-\frac{d^2{\rm ln}{R}}{dt^2}+\frac{2\epsilon}{R^2}
\right)^{-1/2},
\end{eqnarray*}
the Euler-Lagrange equations corresponding to the variational problem \cite{6}
assume the form of a pair of equations for $F(t)$ and~$R(t)$:
\begin{eqnarray}
\label{e7} &&-\frac{d^2{\rm ln}{R}}{dt^2}+\frac{2\epsilon}{R^2}=
\frac{1}{F^2},\\ \label{e8} && \frac{d^2F}{dt^2}+\frac{2\epsilon}{R^2}F=0.
\end{eqnarray}

Exact solutions of this system are the simplest to find for the case
$\epsilon = 0$, which corresponds to a Friedmann flat-space universe.  In
this case we deduce the following from Eqs.\,\,{\ref{e7}) and~(\ref{e8}):
\begin{eqnarray*}
&& -\frac{d^2{\rm ln}{R}}{dt^2}=\frac{1}{F^2}, \quad \frac{d^2F}{dt^2}=0.
\end{eqnarray*}
The solution of these equations is given by the functions
\begin{eqnarray}
\label{e9} R(t)=k_0(a_0t+b_0)^{{1}/{a_0^2}}e^{c_0t}, \quad F(t)=a_0t+b_0,
\end{eqnarray}
where $a_{0}$, $b_{0}$, $c_{0}$, and $k_{0}$ are arbitrary constants.
Substituting the solution (\ref{e9}) into Eqs.\,\,(\ref{e4}) and \cite{5}, we
obtain the following results for the field and the potential:
\begin{eqnarray}
\label{e10} &&\phi(t)=\sqrt{\frac{2}{\kappa}}\,\frac{1}{a_0}{\rm
ln}(a_0t+b_0)+\phi_0,\\
 \label{e11} &&V(\phi, c_0)=\frac{1}{\kappa}\left[\Lambda+3c_0^2+ \left(
\frac{3}{a_0^2}-1\right)e^{-2\alpha(\phi-\phi_0)}+ \frac{6c_0}{a_0}e^{-
\alpha(\phi-\phi_0)}\right],
\end{eqnarray}
where $\alpha = a_{0}\sqrt{\kappa /2}\,$.  It is interesting to note that
the solution (\ref{e9})--(\ref{e11}) generalizes a solution obtained
previously\cite{13} using the method of generation of new solutions by means of
invariant transformations of $\phi $, {\it R}, and {\it V}, which do not
alter the equations of the standard inflation model.  As should be
expected, the given solution describes an inflationary regime of the
exponential or power-law type or a combination thereof.

The solution (\ref{e10}), (\ref{e11}) can be associated with the equation
of state of matter on the basis of the analogy between self-inter\-acting
scalar field theory and an ideal fluid.  This analogy, in particular, is
expressed in the fact that by comparing the energy-momen\-tum tensors of
these two models one can formally calculate the pressure {\it p} and the
energy density $\rho $ of an ideal fluid in terms of the parameters of the
scalar field model according to the rule
\begin{eqnarray}
\label{e12} &&T_4^4=\rho=\frac{1}{2} \dot{\phi}^2+V(\phi),\\
 \label{e13} &&-T_1^1=-T_2^2=-T_3^3=p=\frac{1}{2} \dot{\phi}^2- V(\phi).
\end{eqnarray}
Substituting the expressions for the self-inter\-action potential $V(\phi
)$ from Eq.\,\,(\ref{e11}) and the field $\phi $ from Eq.\,\,(\ref{e10}),
we obtain the relations
\begin{eqnarray*}
&&\rho =\frac{1}{\kappa}\left[\Lambda+3c_0^2+\frac{3}{a_0^2}e^{-2\alpha(
\phi-\phi_0)}+ \frac{6c_0}{a_0}e^{-\alpha(\phi-\phi_0)}\right],\\
 &&p =- \frac{1}{\kappa}\left[\Lambda+3c_0^2+(\frac{3}{a_0^2}-2)e^{-2
\alpha(\phi-\phi_0)}+ \frac{6c_0}{a_0}e^{-\alpha(\phi- \phi_0)}\right].
\end{eqnarray*}
Eliminating the field $\phi $ from these equations, we obtain an effective
equation of state of matter in the form
\begin{eqnarray}
\label{e14} p=-\rho+\frac{2a_0^2}{\kappa}\left(-c_0+\sqrt{ \frac{1}{3}(
\kappa\rho-\Lambda)}\, \right)^2.
\end{eqnarray}

When the class of solutions (\ref{e9})--(\ref{e11}) is extrapolated to
small and large times, i.e., beyond the limits of the inflationary phase
$t_{i} \le t_{1} < t_{2} \le t_{f}$, the following
characteristics are readily established.  The given solution always begins
from a singular state, i.e., for any model parameters ($\Lambda ,\,
a_{0},\, c_{0},\, b_{0}$) there is a time $t_{0} = -b_{0}/a_{0}$ in the
history of the evolution of the scale factor when $R(t_{0}) = 0$, after
which the scale factor increases at an accelerated rate for some time.  A
natural transition to Friedmann expansion does not take place at larger
times, and this is a drawback of the analyzed solutions.  The transition to
the Friedmann regime requires that $1/a_{0}^{2} = 2/3$ and $c_{0} = 0$.  It
is therefore obvious that without the cosmological constant $\Lambda $ in
Eq.\,\,(\ref{e4}) its role is assumed by $c_{0}$, and the condition $c_{0}
= 0$ corresponds to the standard departure from the inflationary
regime.\cite{linde90}

Again we emphasize that the resulting analytical solution
(\ref{e9})--(\ref{e11}) corresponds to an exactly tuned potential in the
slow-roll regime in the variational formulation.

It is a well-known fact that the pioneering work on inflation (see the
surveys in \cite{linde90,5}, and \cite{6}) treated a truncated
system of Einstein and scalar field equations, with the slow-roll regime
defined in such a way as to eliminate the second time derivative of the
scalar field $\ddot{\phi }$ from the equations, a device justified by the
condition $\dot{\phi }^{2} \ll V(\phi )$.  There are no significant
constraints on the self-inter\-action potential in this approach.  In our
case we have obtained an exact profile of the potential in the explicit
form (\ref{e11}).  We note once again that a solution analogous to
(\ref{e11}) has been obtained previously,\cite{13} but without any mention of
the slow-roll regime {\it per se}.  It is easily verified that the
condition $\dot{\phi }^{2} \ll V(\phi )$ is satisfied in a certain time
period $t_{i} \le \Delta t < t_{f}$ for the exact solution
(\ref{e9})--(\ref{e11}).

%\setbox0=\hbox{$\pmb{$V = V(t)$}$}
\section[]{Generation of exact solutions for a given potential
V = V(t)%\box0
}\label{s3}

The problem of analyzing the interrelationship between the scenario of
evolution of the universe and the form of the self-inter\-action potential
can be treated in time scales greater than the inflationary phase.  In this
section, with a view toward completeness, we discuss the overall evolution
of the universe.

In the standard approach the Einstein equations contain the
self-inter\-action potential of the scalar field in the form of the
function $V = V(\phi )$.  In our approach $V(\phi )$ is actually replaced
by the function $V(t)$, which must be interpreted as an effective potential
of material fields.  We note that in the given case of a homogeneous and
isotropic universe the representation of the self-inter\-action potential
in the form $V = V(t)$ does not conflict with the general variational
problem of the derivation of the Einstein equations, where significant use
is made of the functional dependence $V(\phi )$, since each function $V(t)$
and $\phi (t)$ corresponds one-to-one with a particular function $V(\phi
)$.  We call the function $V(t)$ the evolution of the potential energy or
the potential history, underscoring the departure of our approach from the
standard approach, in which the dependence $V = V(\phi )$ is fixed.

We consider the problem of finding all possible types of evolution of the
universe for a fixed potential history $V = V(t)$ describing the time
variation of the potential energy of material fields.  Stated in the
indicated form, the problem has exact solutions, whose construction reduces
to the integration of linear equations in the case $\epsilon = 0$.  We
write Eq.\,\,(\ref{e4}) as an equation for the function $Z(t) = Z^{3}$:
\begin{eqnarray}
\label{e15} -\ddot{Z}+3(\kappa V(t)-\Lambda)Z-6\epsilon Z^{1/3}=0.
\end{eqnarray}

In the case of a flat-space universe ($\epsilon = 0$) this equation
acquires the form of an ordinary linear differential equation
\begin{eqnarray}
\label{e16} -\ddot{Z}+3(\kappa V(t)-\Lambda)Z=0,
\end{eqnarray}
which has the same form as the Schr\"{o}dinger equation for the motion of a
quantum particle in one-dimensional space with a particle potential energy
$U(t) = 3\kappa V(t)$ and self-energy $E = 3\Lambda $.  Now the function
$Z(t)$ assumes the role of the particle wave function.  We wish to examine
this case in more detail.

We assume that one of the solutions of this equation for a fixed dependence
$V = V(t)$ has been found.  We denote it by $Z_{1}(t)$, whereupon a second
linearly independent solution of this equation $Z_{2}(t)$ for a fixed
function $U(t)$ is easily found, because any two linearly independent
solutions of Eq.\,\,(\ref{e16}) are related by the equation,
\begin{eqnarray}
\label{e17} Z_1\dot{Z}_2-Z_2\dot{Z}_1=W_0,
\end{eqnarray}
where $W_{0}$ is a constant.  From this result we obtain
\begin{eqnarray}
\label{e18} Z_2=Z_1\left(Q_0+W_0\int\frac{dt}{Z_1^2}\right),
\end{eqnarray}
where $Q_{0}$ is a constant of integration.  All solutions of
Eq.\,\,(\ref{e16}) corresponding to a specified fixed potential $U(t)$ can
be found by varying the constants $Q_{0}$ and~$W_{0}$.

For example, the solution (\ref{e9}) for $R(t)$ in the problem with a
minimally varying field leads to the solutions
\begin{eqnarray}
R_2(t)&=&R_0(t)\left[Q_0+W_0\int R^{-6}(t){dt} \right]^{1/3} \nonumber \\
 &=&k_0(a_0t+b_0)^{1/a_0^2}e^{c_0t}\left[Q_0+ k_0^6W_0
\int{dt}{(a_0t+b_0)^{-6/a_0^2}e^{-6c_0t}}\right]^{1/3}. \label{e19}
\end{eqnarray}
In the case of pure power-law inflation $C_{0} = 0$ in the solution
(\ref{e9}) with the minimally varying scalar field we obtain a general
class of solutions for $V(t)$ fixed in~(\ref{e10}):
\begin{eqnarray}
R_2^{(p)}(t)&=& k_0(a_0t+b_0)^{1/a_0^2}\left[Q_0+k_0^6W_0
\int{dt}{(a_0t+b_0)^{-6/a_0^2}}\right]^{1/3}= \nonumber\\
 &=& k_0(a_0t+b_0)^{1/a_0^2}\left[Q_0+k_0^6W_0A{(a_0t+b_0)^{-6/a_0^2+1 }}
\right]^{1/3}, \label{e20}
\end{eqnarray}
where $A = a_{0}^{2}/(a_{0}^{2} - 6)$.

Note that the solutions (\ref{e19}) and (\ref{e20}) are new exact solutions
for a parametric dependence $V = V(\phi )$, i.e., $V = V(t)$, $\phi = \phi
(t)$, indicated in the solution (\ref{e10})--(\ref{e11}).

The proposed method based on (\ref{e17}) carries over without too much
difficulty to the case $\epsilon \neq 0$.  It is necessary here to fix the
function
\begin{eqnarray*}
Q(t)=3(\kappa V(t)-\Lambda)-6\epsilon Z^{-2/3}
\end{eqnarray*}
as a time function.  The potential $V(t)$ then depends on the form of the
solution.  The case $\epsilon \neq 0$ requires separate investigation.

The linearity of Eq.\,\,(\ref{e16}) for the function $Z(t)$ permits the
formulation of a problem in eigenfunctions and eigenvalues, the role of
which is taken here by the cosmological constant, provided that the given
equation is augmented with homogeneous initial conditions.  For example, we
can investigate all possible evolutions for a fixed function $V(t)$ such
that evolution begins at a certain time $t_{0}$ from the state $R(t) = Z(t)
= 0$ and returns to the same singular state at another time $t_{1} >
t_{0}$.  Such scenarios of the evolution of the universe can be set in
correspondence with oscillating solutions \cite{weinberg75}:  The universe
originated and disappeared in the time period $(t_{0},\, t_{1})$.  The corresponding
problem appears as follows:
\begin{eqnarray}
\label{e21} &&-\ddot{Z}+3(\kappa V(t)-\Lambda)Z=0,\\
 \label{e22} &&Z|_{t=t_0}=0,\quad Z|_{t=t_1}=0.
\end{eqnarray}
This problem has the same form as quantum-mechan\-ical problems for a
discrete spectrum.  If the initial conditions are set at times $t_{0} = -
\infty $ and $t_{1} = +\infty $, and the potential energy is a smooth
function of time, the universe can evolve from other than a singular state
and again evolve into a nonsingular state in the sense that the energy
densities and other physical characteristics of matter (field) are finite.
Other types of homogeneous initial conditions are also possible.  The fact
that only a finite or denumerable set of admissible values of the
cosmological constant occurs for each problem of this kind can shed light
on the issue of the true value of the cosmological constant and the
physical reasons for it.

\section[]{Analysis of the evolution of the universe for various types of
potentials}\label{s4}

The determination of the special characteristics of the origin and
evolution of various inflationary regimes can be pursued on the basis of
the representation (\ref{e16}) in the example of a series of models that
are simple from the standpoint of the construction of solutions but are
interesting from the standpoint of the behavior of the potentials.  In the
present study we have chosen the following model potentials in this
category:
\begin{eqnarray}
&&3\kappa V(t)=2t^2, \label{ea}\\ %\eqnum{a}  \\
 &&3\kappa V(t)=\frac{m}{ t^2}, \quad m={\rm {\rm const}}, %\eqnum{b}
\label{eb} \\
 &&3\kappa V(t)=-\frac{2\lambda_0}{\cosh ^2(\lambda_0 t)}. %\eqnum{c}
\label{ec}
\end{eqnarray}

We investigate the solutions for the potentials (\ref{ea}), (\ref{eb}), and
(\ref{ec}) from the viewpoint of the possible existence of inflationary
regimes and their transition to a Friedmann phase of evolution.

(a) It is readily verified that the potential (\ref{ea}) admits oscillating
solutions corresponding to the statement of the problem in the form
(\ref{e21}), i.e., laws of the evolution of the scale factor such that the
universe goes from a singularity in the limit $t \to -\infty $ to a new
singularity in the limit $t \to +\infty $.  Thus, the solutions of
Eq.\,\,(\ref{e16}), subject to null boundary conditions as $t \to \pm
\infty $, have the form
\begin{eqnarray*}
Z(t)=H_i(t)e^{-t^2/2},
\end{eqnarray*}
where $H_{i}(t)$ are Hermite polynomials.  For example, the simplest
solution has the form $Z(t) = d_{0}^{3}\exp \{-t^{2}/2\}$.  In this case
\begin{eqnarray*}
R(t)=d_0\exp\{-t^2/6\}, \quad \phi(t)=\pm \sqrt{\frac{2}{3\kappa}}\,t+
\phi_0, \quad V(\phi)= (\phi(t)- \phi_0)^2.
\end{eqnarray*}
The value of the cosmological constant ensuring this evolution regime is
$\Lambda = 1/3$ (in appropriate units of measurement for the given
problem).  For this particular solution the condition
\begin{eqnarray*}
\ddot R=(d_0/3)\exp\{-t^2/6\}\{{t^2}/{3}-1\}>0
\end{eqnarray*}
implies the start of inflation at $t > \sqrt{3}\,$ and no exit from the
inflationary regime.

Other Hermite polynomials correspond to higher absolute values of $\Lambda
= 1,\, 5/3,\dots \,$.  The evolution regimes in this case are such that the
universe passes through a singular state several times.  It is readily
verified that the inflationary phase for these oscillating solutions begins
at $t \to -\infty $, where the scale factor has a zero minimum, and ends at
a time $t_{0}$ corresponding to an inflection point of the function $R =
R(t)$.

(b) Potentials of the type (\ref{eb}) for $m \ge =1/4$ are
interesting because in the case $\Lambda = 0$ they describe all possible
types of power-law evolution of the scale factor, including power-law
inflation.  In fact, the general solution of Eq.\,\,(\ref{e21}) for
arbitrary $m > 0$ and $\Lambda = 0$ has the form
\begin{eqnarray}
\label{e23} Z(t)=C_1 t^{\alpha}+ C_2 t^{\beta},
\end{eqnarray}
where $\alpha $ and $\beta $ are distinct solutions of the algebraic
equation $x^{2} - x - m = 0$, i.e.,
\begin{eqnarray*}
\alpha=\frac{1}{2}+\sqrt{\frac{1}{4}+m}>0,\quad \beta=\frac{1}{2}-\sqrt{
\frac{1}{4}+m}<0.
\end{eqnarray*}
It is evident from this result that in the case $C_{1} > 0$, $C_{2} > 0$
the solution (\ref{e23}) is positive at $t > 0$ and has one minimum at a
point $t_{0} > 0$; after passage through this minimum, power-law inflation
begins at a time $t_{1} > t_{0}$, asymptotically approaching the $t^{\alpha
/3}$ regime.  The asymptotic power-law inflation regime corresponds to the
requirement $\alpha > 3$, so that $m > 6$.  We note that prior to the time
$t_{1}$ the field $\phi $ is imaginary, and the evolution regime physically
achieved in the self-inter\-acting scalar field model begins precisely at
$t_{1}$.  To circumvent this problem, it must be required that the
conditions $C_{1} > 0$ and $C_{2} \le 0$ hold.  In this case
evolution begins from a singular state at a time $t_{S}$ and makes an
immediate transition to power-law inflation.

Remarkably, the only regime of power-law evolution of the scale factor that
does not lead to the potential (\ref{eb}) is Friedmann expansion with the
ultimately rigid state of matter $p = \rho $, for which $R(t) \propto
t^{1/3}$, so that $Z \propto t$ and $V(t) = -0$ for $\Lambda = 0$.  The
Friedmann expansion regime $R(t) \propto t^{1/3}$ sets in as $t \to \infty
$ subject to the condition that as the potential decreases, it tends to
zero more rapidly than $1/t^{2}$.  An example of this kind of behavior is
afforded by one of the possible functional dependences $R = R(t)$, based on
evolution of the form
\begin{eqnarray*}
Z(t)=z_0+\frac{t^2}{1+t}.
\end{eqnarray*}
At $t > -1$ the scale factor passes through a minimum value $z_{0}^{1/3}$
at the time $t = 0$, after which inflation begins.  Inflation ends when the
inflection point is reached, and then as $t \to +\infty $, the evolution
asymptotically settles into the Friedmann regime $R(t) \propto t^{1/3}$.
The self-inter\-action potential in this case is
\begin{eqnarray*}
V(t)=\frac{\Lambda}{\kappa}+\frac{2}{3\kappa(1+t)^2}\,\frac{1}{t^ 2+z_0 t
+z_0},
\end{eqnarray*}
whence it follows that
\begin{eqnarray*}
V(t)\to \frac{2}{3\kappa}\,\frac{1}{t^4}\quad {\rm as} \quad t\to\infty.
\end{eqnarray*}

For transition to the Friedmann regime with dustlike matter, i.e., when $p
= 0$, it is sufficient that $V(t) \to 2/t^{2}$ as $t \to \infty $.  For any
other equation of state of the type $p = \gamma \rho $, $\gamma =
{\rm const}$, it is sufficient that $V(t) \to m/t^{2}$, $m \ge -1/4$
as $t \to \infty $.  The case $m = -1/4$ corresponds to a
radi\-ation-domi\-nated state of matter.

The potential (\ref{eb}) also leads to oscillating solutions for $m < 0$,
which correspond to $\Lambda < 0$.  For example, the general solution of
Eq.\,\,(\ref{e16}) for the potential (\ref{eb}) in the case $m = -2$,
$\Lambda < 0$ has the form
\begin{eqnarray}
\label{e24} Z(t)=A \left(k-\frac{1}{t} \right)e^{k t}+B \left(-k-
\frac{1}{t}\right)e^{-k t},\quad k=\sqrt{-3\Lambda}.
\end{eqnarray}
The set of solutions of Eq.\,\,(\ref{e24}) includes one that satisfies the
condition $Z(\pm \infty ) = 0$.  It is the solution corresponding to
$\Lambda = 0$ and has the form $Z = A/|t|$, $A > 0$.

(c) To discern certain characteristics of the limitations imposed on the
growth rate by the condition for transition to Friedmann expansion, we give
additional consideration to the history of the potential (\ref{ec}).  Like
the potential (\ref{ea}), it increases as $t \to \infty $, but only to a
finite value equal to zero.  In this case the solution for $Z(t)$ with
$\Lambda < 0$ has the form
\begin{eqnarray*}
Z(t)=A(\lambda - \lambda_0\tanh (\lambda_0 t))e^{\lambda t} + B(\lambda +
\lambda_0\tanh (\lambda_0 t))e^{-\lambda t}.
\end{eqnarray*}
Here $\lambda ^{2} = -3\Lambda > 0$.  For $\lambda = \lambda _{0}$ this
potential corresponds to an oscillating evolution of the form
\begin{eqnarray*}
Z(t)=\frac{C}{\cosh(\lambda_0 t)}, \quad R(t)=C^{1/3}\cosh^{-1/3}(
\lambda_0 t),
\end{eqnarray*}
where {\it C} is an arbitrary constant.  This is a unique solution for the
given potential at $\lambda = \lambda _{0}$, corresponding to a unique
bound state.  As in the case of the potential (\ref{ea}), it describes an
inflationary regime in the interval $(-\infty ,\, t_{0})$, where $t_{0}$ is
the inflection point of the function~$R(t)$.

In the case $\lambda = \Lambda = 0$ the solution is the function
\begin{eqnarray*}
R(t)=C\tanh^{1/3}\lambda_0 t,
\end{eqnarray*}
which describes the progress of evolution from a singular state at the time
$t = 0$ to its asymptotic transition to the stationary state $R = C =
{\rm const}$ as $t \to +\infty $.

In the case $\lambda > \lambda _{0}$, $A > 0$, $B = 0$ we have the solution
\begin{eqnarray*}
Z(t)=A(\lambda-\lambda_0\tanh(\lambda_0 t))e^{\lambda t}, \quad
R(t)=A^{1/3}e^{\lambda t/3}\left(\lambda-\lambda_0\tanh(\lambda_0 t)
\right)^{1/3},
\end{eqnarray*}
which describes evolution without singularities and with the onset of the
inflationary regime beginning at a time $t_{0} > 0$.

The solutions corresponding to $\Lambda > 0$ are oscillating solutions with
multiple passages through the value $Z = 0$.  It is evident that asymptotic
Friedmann expansion does not occur for this potential.

Comparing the solutions obtained for the three potential types, we arrive
at the following conclusions.  First, potentials that increase as $t \to
\infty $ [potentials (\ref{ea}) and (\ref{ec})] do not allow any transition
to the Friedmann regime if the growth rate of the potential energy exceeds
the rate at which the function $mt^{-2}$  with $m = -1/4$ approaches zero.
Transition to the Friedmann regime requires that the potential decrease
with time according to a power law of the type $mt^{-2}$ with $m > 0$ or
that it increase by an analogous law with $-1/4 < m < 0$.  Second, all
three potential types exhibit the existence of the inflationary regime,
corroborating the conclusion that this regime is not selective with regard
to the form of the potential.  Altogether these considerations indicate
that the most preferred models are those of the type (\ref{eb}), which
require slight modifications to ensure transition to the required Friedmann
regime.  For example, one such modification is the introduction of a weak
dependence of {\it m} in the equation for the potential (\ref{eb}) on the
temperature {\it T} of matter
\begin{eqnarray}
\label{e25} V(t,T)=\frac{m(T)}{t^2}
\end{eqnarray}
such that $m(T) > 0$ near the minimum of $R(t)$ and that $m(T) \to 2$ in
the limit $t \to \infty $.  For example, the required dependence $m(T)$ can
be one with jumps characterizing phase transitions in early universe
matter.  The potential (\ref{e11}) falls in this category in the case
$c_{0} = 0$.  Consequently, these potentials (\ref{eb}) satisfy the
slow-roll principle in the variational formulation set forth in
Sec.\,\,\ref{s2}.

It should also be noted that the simplicity of the potential equation
(\ref{e25}) is preserved only when it is written in the form $V = V(t)$.
If the form of the function $V = V(\phi ,\, T)$ is written on the basis of
the solutions obtained for $R(t)$ and $\phi (t)$, this function acquires an
intricate form and depends significantly on the constants of integration of
Eq.\,\,(\ref{e16}), which, in turn, are determined by the initial
conditions for $R(t)$ and $\phi (t)$.  This consideration demonstrates the
preferability of the representation $V = V(t)$ over $V = V(\phi )$ for
analysis of the model dynamics.

\section{The exact and approximate solutions of inflationary models}

The standard cosmological model of SSF, including an inflationary stage,
in the case of spatially-flat homogeneous and isotropic universe is
described by the following system of equations \cite{linde90}
\beqa{EqPrim}
 &&H^2=\frac{\gvk}{3}\left(\frac{1}{2}{\dot{\phi}}^2+V(\phi)\right),\\
 \label{EqPrim1}
 &&{\ddot{\phi}}+3H{\dot{\phi}}=-\frac{d}{d\phi}V(\phi)
\eeqa
where $\phi$ is a scalar field, $V(\phi)$ is a self-interaction potential
of the scalar field,
$$
  H(t)=\Dt\lln R(t),
$$
$R(t)$ is a scalar factor, $\gvk$ is a gravitational constant.
The usual process of construction of approximate solutions for inflationary
regim consist of neglecting of the values $\ddot\phi$ and $\dot\phi^2$
in comparision with $H\dot\phi$ and $H^2$.
This is the basis of the slow roll approximation.
As a result of such procedure the equations above reduced to the
following simple form
\beqa{EqCut}
 &&H^2=\frac{\gvk}{3}V_с(\phi),\\
 \label{EqCut1}
 &&3H{\dot{\phi}}=-\frac{d}{d\phi}V_с(\phi),
\eeqa
in which thier integration is a more simple task.
For example, follow by the review \cite{gmss92}, one can obtain
\beqa{SolR}
      &&R_с(\phi)=R_0\exp\{-\gvk\int\frac{V}{V'}d\phi\}
\eeqa
For the potential in the trancased equations
\rf{EqCut}-\rf{EqCut1} and for thiers solutions
\rf{SolR} the index ``с'' will be used.

We prove here that the exact equations \rf{EqPrim}-\rf{EqPrim1} can be
presented in the form of approximate ones
\rf{EqCut}-\rf{EqCut1} with the help of an effective potential
of self-interaction introduced in consideration.
We have the formal relation % Формально справедливо следующее соотношение
\beq{dpU}
     \dot{\phi}=U(\phi)
\eeq
where $U(\phi)$ is a function of $\phi$. The relation \rf{dpU} is true
for any dependence $\phi=\phi(t)$.
Let us introduce now an effective potential $W(\phi)$ following by the rule
\beq{ConVV}
      W(\phi)=V(\phi)+\frac{1}{2}U^2(\phi).
\eeq
Function $W(\phi)$ represents the total energy of the feild as the function
of its value.
Using \rf{ConVV} the equations
\rf{EqPrim}-\rf{EqPrim1} can be transfermed to
\beqa{EqSec}
 &&H^2=\frac{\gvk}{3}W(\phi),\\
 \label{EqSec1}
 &&3H{\dot{\phi}}=-\frac{d}{d\phi}W(\phi),
\eeqa
As easily seen, with the accuracy of the exchange
of $V_с(\phi)$ on $W(\phi)$ equations
\rf{EqSec},\rf{EqSec1} are equivalent to trancased equations \rf{EqCut},
\rf{EqCut1}. A scalar factor \rf{SolR} in the exact model is
\beqa{SolRexcp}
   &&R(\phi)=R_0\exp\{-\gvk\int\frac{W}{W'}d\phi\}
\eeqa
or
\beqa{EqRt}
  &&R(t)=R_0\exp\{-\int\limits_{t_0}^t\sqrt{\frac{\gvk}{3} W(t)}dt\}
\eeqa
The equations of SSF in the form \rf{EqSec}-\rf{EqSec1} are the starting point
of the model under consideration admitting an exit to the friedmann stage
of evolution.

Equations  \rf{EqSec},\rf{EqSec1} lead to the relation
\beq{ConVVU}
      \sqrt{3\gvk}UW^{1/2}=-W'
\eeq
which along with \rf{ConVV} describe the relation between functions
$W(\phi)$ and $U(\phi)$ in a such a way that just only one of the functions
above is arbitrary.
By solving this equation respective to $W$ one can obtain
\beq{SolWU}
 W(\phi)=\frac{3\gvk}{4}\left(\int U(\phi)d\phi+\sqrt{W_0}\right)^2\ge 0,
\eeq
where $\sqrt{W_0}$ is a real constant of integration.
This means that the total energy and, consequently, the effective
potential $W$ is of nonnegative value in the model under consideration.

In future we will use the following terminology for the potentials
$V(\phi)$ и $W(\phi)$.
The potential of SSF ССП $V(\phi)$ presenting in exact equations
\rf{EqPrim}-\rf{EqPrim1} will call {\it true or physical potential}
The effective potential $W(\phi)$ will call
{\it effective potential of the total energy} (for the sake of brivity
-- the potential of total energy)
with the aim to differ it from the effective potentials widely used
in quantum field theories.

\section{The model with effective Higgs potential} \label{phi4}

In order to describe an exit from inflation to Freidmann's regim under
condition $t\to\infty$ we have to formulate an asimptotical conditions of
such exit. The sufficient condition following from the observational data
is the vanishing of the energy of an inflaton field when $t\to\infty$.
Thus we can require the asimptotical condition
\beq{AsECon}
    \frac{1}{2}(\dot\phi(t))^2+V(\phi)\to 0, ~~{\rm при} ~~t\to\infty.
\eeq
The last can be decomposed for two another conditions
\beqa{AsPCon}
 &&\dot\phi(t))=U(\phi(t))\to 0, ~~{\rm при} ~~t\to\infty,\\
 \label{AsVCon}
 &&V(\phi(t))\to  0, ~~{\rm при} ~~t\to\infty.
\eeqa

One of the simplest models satisfing by the requirement \rf{AsPCon}
is the model in which
\beq{Ui}
     \dot{\phi}=U(\phi(t))=u_0 e^{-\mu t}, ~~~\mu>0.
\eeq
For \rf{Ui} we have
\beq{Pi}
      \phi(t) = \phi_\infty  -  \frac{u_0}{\mu} e^{-\mu t}.
\eeq
Therefore
$$
     U(\phi)=\mu(\phi_\infty-\phi(t)).
$$
Inserting the last relation to \rf{ConVVU}, one can find
\beqa{Vefi}
&&W(\phi)=\left(\frac{\gl}{4}(\phi-\phi_\infty)^2-\sqrt{V_\infty}\right)^2,\\
\label{Vi}
&& R(\phi)=R_0\exp\left\{-\gvk\int\limits_{\phi_0}^\phi\frac{\gl(\phi-\phi_\infty)^2-4\sqrt{V_\infty}}{
4\gl(\phi-\phi_\infty)}d\phi\right\}=\\ \nonumber
&&= R_0\exp\left\{\frac{\gvk}{8}(\phi_0-\phi_\infty)^2\right\}
\left(\left|\frac{\phi-\phi_\infty}{\phi_0-\phi_\infty}\right|\right)^{\ga}
\exp\{-\frac{\gvk}{8}(\phi-\phi_\infty)^2\}
\eeqa
Here
$$
\gl=\mu\sqrt{3\gvk}, ~~~\ga=\frac{\gvk}{\gl}\sqrt{V_\infty}.
$$
${V_\infty}$ is the value of the potential in the limit
$t\to\infty$, and $\phi_0$ is the value of the field at the starting moment
$t_0$. It is easily to check that for the potential \rf{Vi}
the condition \rf{AsVCon} is valid for the case if $V_{\infty}=0$.

From the relation \rf{ConVV} one can find the physical potential
\beq{Vhiggs}
V(\phi)=\left(\frac{\gl}{4}(\phi-\phi_\infty)^2-\sqrt{V_\infty}\right)^2-
\frac{1}{2}\mu^2(\phi_\infty-\phi)^2.\\
\eeq
This is well known potential of $\phi^4$-theory with Higgs properties of
spontaneous breaking of the symmetry. In accordance with this theory
the effective mass of Higgs boson will be equal to
$$
 M_h=\sqrt{\gl\sqrt{V_\infty}+\mu^2}.
$$
The minimima of the potential $V(\phi)$ are located in the points
$$
\phi^{\pm}=\phi_\infty\pm2\sqrt{\gl\sqrt{V_\infty}+\frac{\mu^2}{\gl^2}}=
 \phi_\infty\pm 2\sqrt{\mu\sqrt{3\gvk V_\infty}+\frac{1}{3\gvk}}
$$
and maximum in the point $\phi_\infty$ (see Fig. 1).
\begin{figure}
\centering
\psfig{file=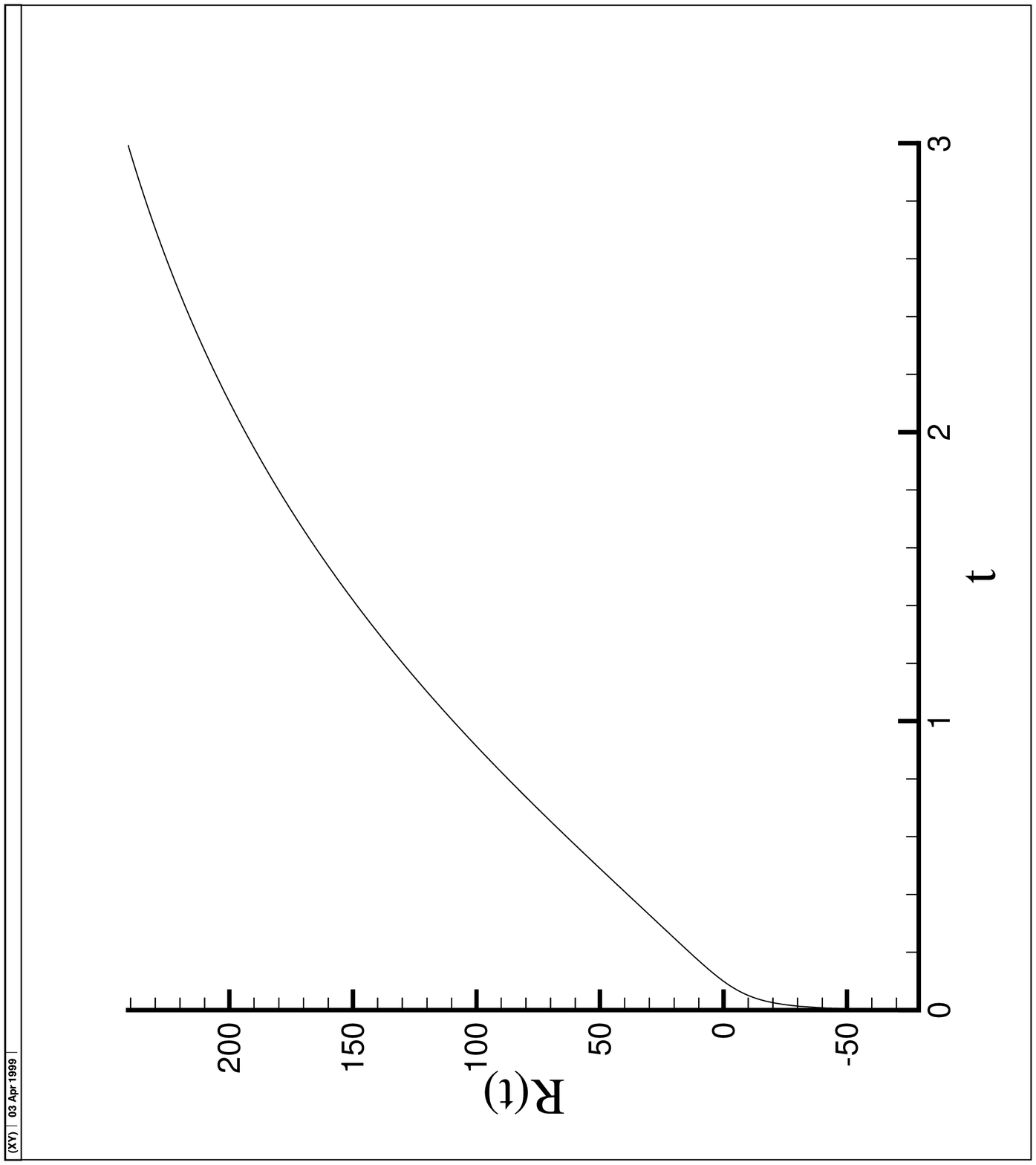,width=8cm}
\caption{ .}\label{fig1}
\end{figure}

In the case $V_{\infty}=0$ we have
\beq{Mh}
   M_h=\mu,~~~\phi^{\pm}= \phi_\infty\pm 2\sqrt{\frac{1}{3\gvk}}.
\eeq

The set of formulas \rf{Vefi}-\rf{Vi}, \rf{Vhiggs} and \rf{Pi}
gives the exact solution for a scale factor of the cosidered model.
By inserting in expression for $R(\phi)$ the dependence $\phi=\phi(t)$,
one can obtain
$$
    R(t)=R_0\exp\left\{\frac{\gvk}{8}(\phi_0-\phi_\infty)^2\right\}\left(-\frac{u_0}{\mu}\right)^{\ga}
    \left(\left|\frac{1}{\phi_0-\phi_\infty}\right|\right)^{\ga}
    e^{-\mu\ga t}
    \exp\left\{-\frac{\gvk u_0^2}{8\mu^2}e^{-2\mu t}\right\}
$$
The condition \rf{AsVCon}, as it was mention, leads to the requirement
$V_\infty=\ga=0$, which give more simple form for scalar factor evolution
\beq{ModInf}
    R(t)=R_\infty
    \exp\left\{-\frac{\gvk u_0^2}{8\mu^2}e^{-2\mu t}\right\}.
\eeq
где
$$
   R_\infty=R_0\exp\left\{\frac{\gvk}{8}(\phi_0-\phi_\infty)^2\right\}.
$$
\begin{figure}
\centering
\psfig{file=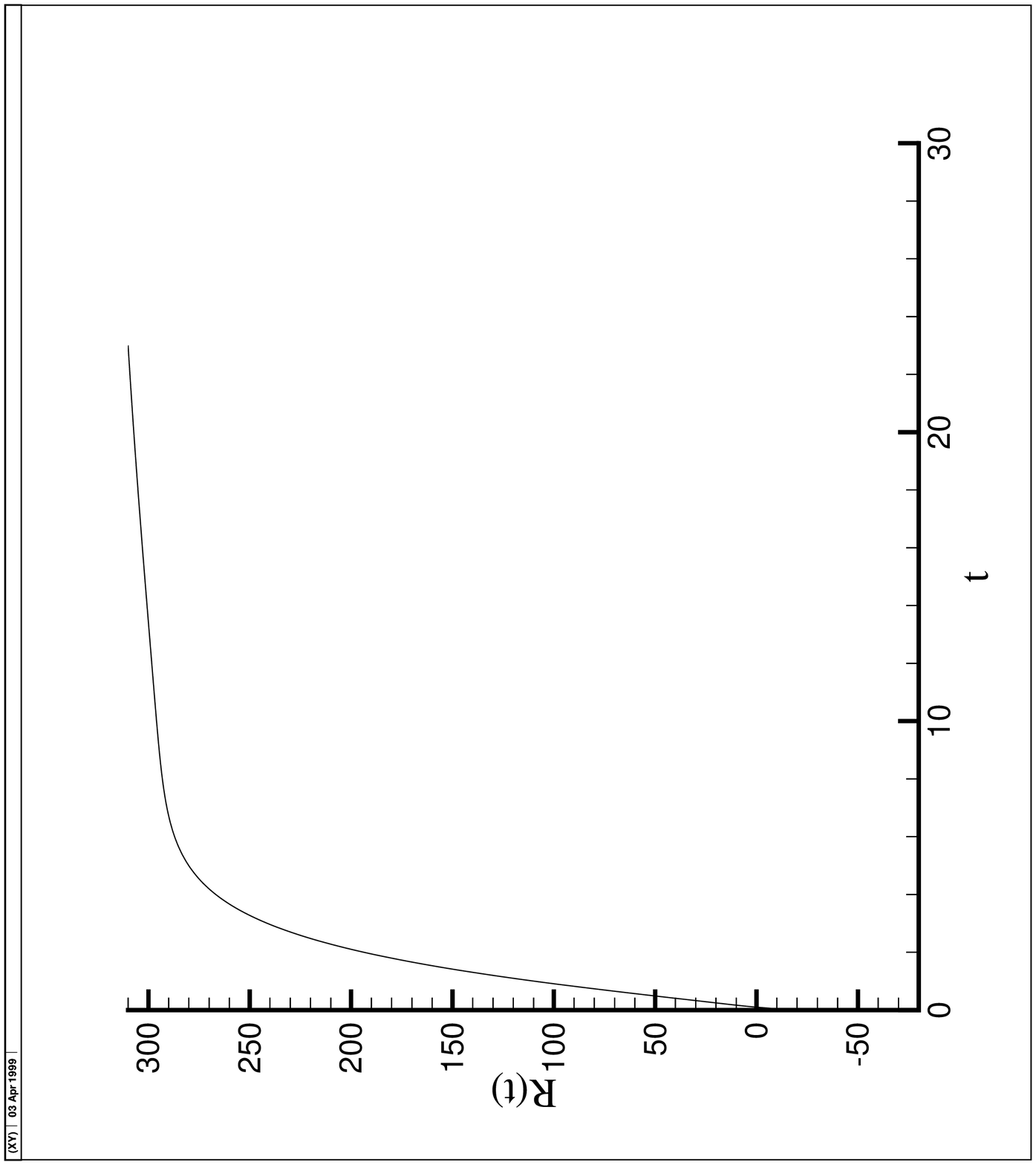,width=8cm}
\caption{ .}\label{fig2}
\end{figure}

This regim of an evolution corresponds to chaotic inflationary model
with a massive scalar fiels \cite{linde90},
besides the initial conditions give the following bounds:
$U(\phi)\approx M_P^2,~~ u_0\approx M_P^2 $,
where $M_P=\sqrt{\hbar c/G}$ is the Plank's mass.

It is follow from the relations \rf{Mh} that the conditions \rf{Ui}
and \rf{Pi} can be considered as a classical equivalent of decay of
Higgs bosons with the mass $M_h=\mu$ and with the characteristic time of
decay  $\tau_h \approx 1/\mu$.
Thus in the given model it is the decay of Higgs bosons that provide
an inflationary temp of expantion. It shoud be mentioned here that
this model contains the exit from inflation but to the asimptotically
static Universe. Therefore this model can be considered as auxiliary
model admitting an important exit property.
To obtain a correct asymptotics when $t\to\infty$ the model above shoud
be modified with the account of the following fact: the transition to
the Freidmann stage is possible if there exists after the moment of Higgs
bosons decay a matter in the universe
in the condition close to thermodynamical equilibrium with the equation
of state $p=\Gg \veps$. In this case an adiabatical expantion of the universe
will be of Freidmann type.
Using the suggestion above we consider the next auxiliary model.

\section{Models with exit to Freidmann's evolution} \label{Fr}

In the work \cite{ZCS98} (see also the section 3.) have been analysed
the evolution's regim's corresponding to the condition
\beq{Ve}
    V(\phi(t))=\frac{m}{3\gvk t^2}
\eeq
where $m$ is the constant, which satisfy to the restriction $m\ge -1/4$.
In this case the scalar factor's evolution is determined by
\beq{Eqt-2}
    R(t)=С t^{\gb/3},
\eeq
where
$$
\gb=\frac{1}{2}+\sqrt{\frac{1}{4}+m}>0.
$$
The equations of states can be obtained from the formula
$$
   p=\Gg \veps, ~~~\Gg=\frac{\gb-m}{\gb+m}=\frac{1-2m+\sqrt{1+4m}}{1+2m+\sqrt{1+4m}}
$$
Generally speaking, the solution \rf{Eqt-2} can be considered with arbitrary
values of $\gb>0$ including a power law inflation under $\gb>3$.
But we restrict ourselves by Freidmann's regims when $0<\gb<3$.
The case of power law inflation has been considered in the works
\cite{ch97gc, czs97, ZCS98}.

The limiting case $m=-1/4$ corresponds to equation of state with
$\Gg=3$, the case $m=0$ -- equation of ultrastiff matter with
$\Gg=1$, the case $m=2$ -- dust matter with $\Gg=0$,
and the case $m=3/4$ -- pure radiation with $\Gg=1/3$.
In general all Freidmann's regim with $\gb\le 1$ correspond to
condition $-1/4\le m \le 6$.
The model under consideration corresponds to adiabatical expantion of the
universe filled by the matter with an equation of state close to equilibrium
one.
The special interest is to the radiation era $p=\frac{\epsilon}{3}$
and matter dominated epoch $p\approx 0$ \cite{zelnov75}.

In the framework of our approach if we have chosen the function
$V(t)$ as \rf{Ve} then
\beq{Ue}
     U(t)=\frac{\nu}{t}
\eeq
Under this functional dependence $U$ on $t$ the conditions \rf{AsPCon} and
\rf{AsVCon} are true as well.
Following by \rf{dpU},\rf{ConVV},\rf{EqSec}, we obtain the relations
\beqn
 &&\phi(t)=\phi_1+\nu\lln t,~~U(\phi)=\nu \exp\{-\frac{1}{\nu}(\phi-\phi_1)\},
 ~~\phi_1=\const,\\
 &&W(\phi)=\left(\sqrt{V_1}+\frac{\sqrt{3\gvk}}{2}\nu^2\exp\{-\frac{1}{\nu}(\phi-\phi_1)\}\right)^2,\\
 &&V(\phi)=\left(\sqrt{V_1}+\frac{\sqrt{3\gvk}}{2}\nu^2\exp\{-\frac{1}{\nu}(\phi-\phi_1)\}\right)^2-
	   \frac{1}{2}\nu^2\exp\{-\frac{2}{\nu}(\phi-\phi_1)\}.
\eeqn
By inserting in expression for $V(\phi)$ the functions $\phi=\phi(t)$
we obtain
$$
   V(t)=\left(\sqrt{V_1}+\frac{\sqrt{3\gvk}}{2}\frac{\nu^2}{t}\right)^2-
           \frac{1}{2}\frac{\nu^2}{t^2}.
$$
The dependence $R=R(t)$ is defined from \rf{EqRt} and for the case under
consideration it has the form
\beq{LREx}
     R(t)=R_0t^{\gvk\nu^2/2}\exp\{\sqrt{\frac{\gvk}{3}V_1} t\}
\eeq
All possible power law scalar factor's evolution corresponding to
the dependence of the potential energy on time
\rf{Ve}, can be obtained in the case $V_1=0$.
It is the case that one has an axit on Freidmann's regims
\rf{Eqt-2}, besides
$$
       \gb=\frac{3}{2}\gvk\nu^2.
$$

We have to mention, that in the model under consideration, based on
\rf{Ve} and \rf{Ue}, an inflation is absent, but Freidmann's regim
will appear asimptotically when
$t\to\infty$ for any other evolution's regim differing from
\rf{Ve} and \rf{Ue} by existing of fast decreasing terms.
It is just necessary that the speed of decreasing is over
$t^{-2}$ in \rf{Ve} and $t^{-1}$ in \rf{Ue}, for example, by an exponential
rate. This point is the basis for the new model generalizing the results
obtained in sections \ref{phi4} and \ref{Fr}.

\section{Inflationary model with the exit to the Freidmann's expantion.}

To construct the model of univers' evolution with inflationary stage
admitting the exit to radiation dominated stage and matter dominated epoch
let us consider the linear superposition of the exialary models,
which have been considered in the sections \ref{phi4} и \ref{Fr}.
Namely, let us consider the model corresponding to the relations
\beqa{Uie}
 &&\dot{\phi}= U(t)=u_0e^{-\mu t}+\frac{\nu}{t},\\
 \label{Pie}
 &&\phi(t)=\phi_c-\frac{u_0}{\mu} e^{-\mu t}+\nu\lln t.
\eeqa
In this model the conditions \rf{AsPCon} are satisfied.
Given model has three parameters the meaning of which are clear from
preveous analysis of the models \rf{Ui} and \rf{Ue}.
An inflation regim is possible if the period in the history
of the evolution exist when
\beq{Conumn}
       u_0e^{-\mu t} >> \frac{\nu}{t}
\eeq
Evidently, we can choose the parameters $\mu,\nu$ and $u_0$
in such a way that the condition \rf{Conumn} will be satisfied.
Let us show this by direct calculations.

It is difficult to obtain a direct dependence for the function  $U=U(\phi)$
using the relations \rf{Uie} and \rf{Pie}.
Nevertheless the presentation in the form of function $U=U(t,\phi)$ is
possible. For example,
$$
   U(\phi,t)=\frac{\nu}{t}+\nu\mu\lln t -\mu(\phi-\phi_c)
$$
In the present form we can control the deviation the model \rf{Uie} from the
model \rf{Ue}.

Using the relations \rf{ConVVU}, \rf{Uie}, \rf{Pie} one can calculate
the effective potential $W$
\beqa{EqVef}
&&W(t)=\left(\sqrt{V_0}+\int\limits_{t_0}^tU^2(t)dt\right)^2=
 \left(-\frac{u_0^2}{2\mu}e^{-2\mu t}-\frac{\nu}{t}+2u_0\nu\int\limits_{t_0}^t\frac{e^{-\mu t}}{t}dt\right)^2,\\
&&V(t)=\frac{3\gvk}{4}\left(\sqrt{V_0}-\frac{u_0^2}{2\mu}e^{-2\mu t}-
\frac{\nu}{t}+2u_0\nu\int\limits_{t_0}^t\frac{e^{-\mu t}}{t}dt\right)^2-
\left(u_0e^{-\mu t}+\frac{\nu}{t}\right)^2
\eeqa
We will investigate the potentials
$W(\phi)$ and $V(\phi)$ as the functions of time, because it is impossible to
find the explisit dependence on $\phi$. For this purpose it is suatable
to use the relation \rf{EqRt}.

The change of scalar factor $R(t)$ versus the time is shown in the Fig.3
and Fig.4 for the models \rf{Uie}, \rf{Pie}.
Fig. 3 corresponds to the period in conventional units before an inflation.
Fig. 4 corresponds to  the case of general evolution of the universe.
\begin{figure}
\centering
\psfig{file=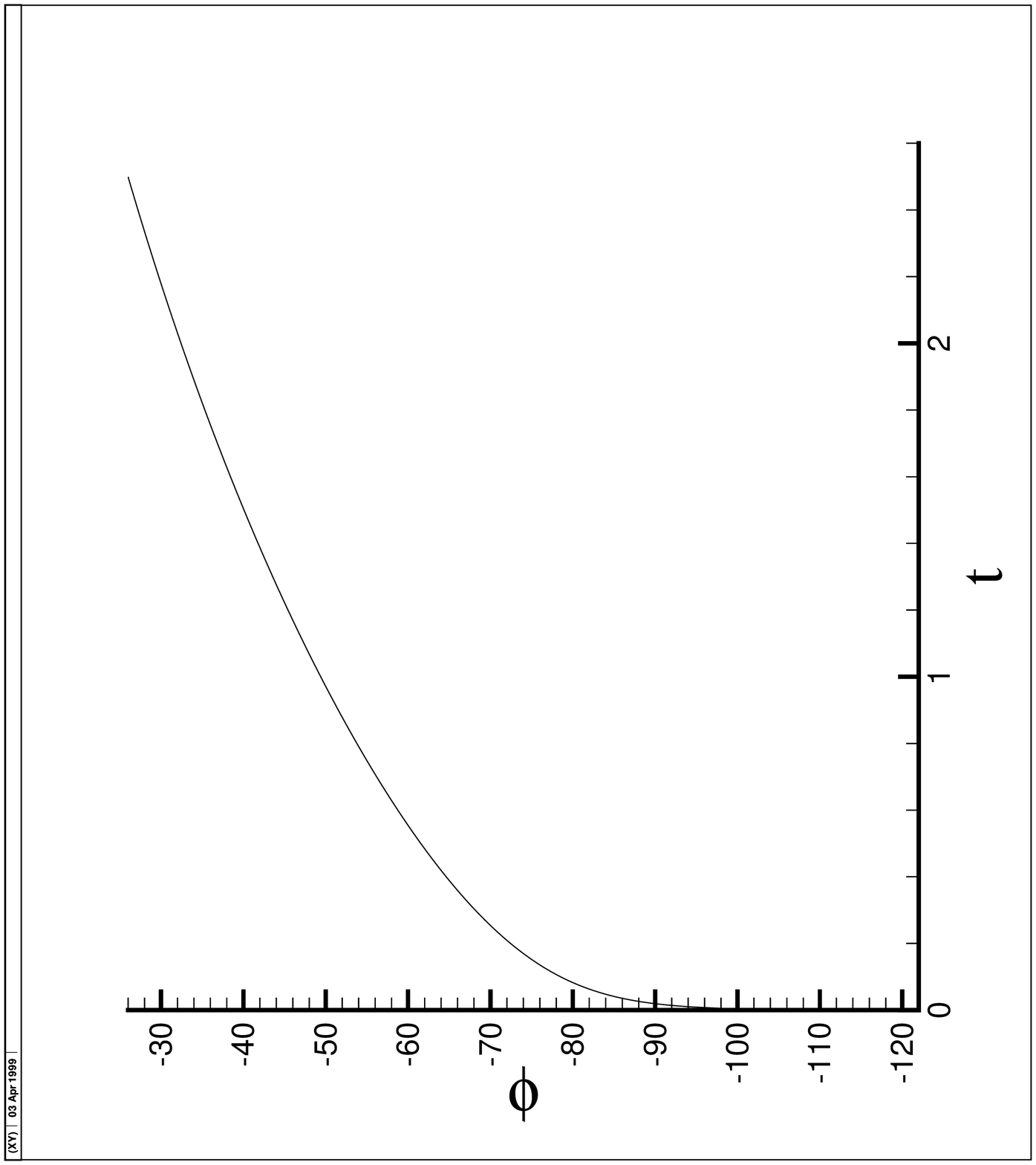,width=8cm}
\caption{ .}\label{fig3}
\end{figure}
\begin{figure}
\centering
\psfig{file=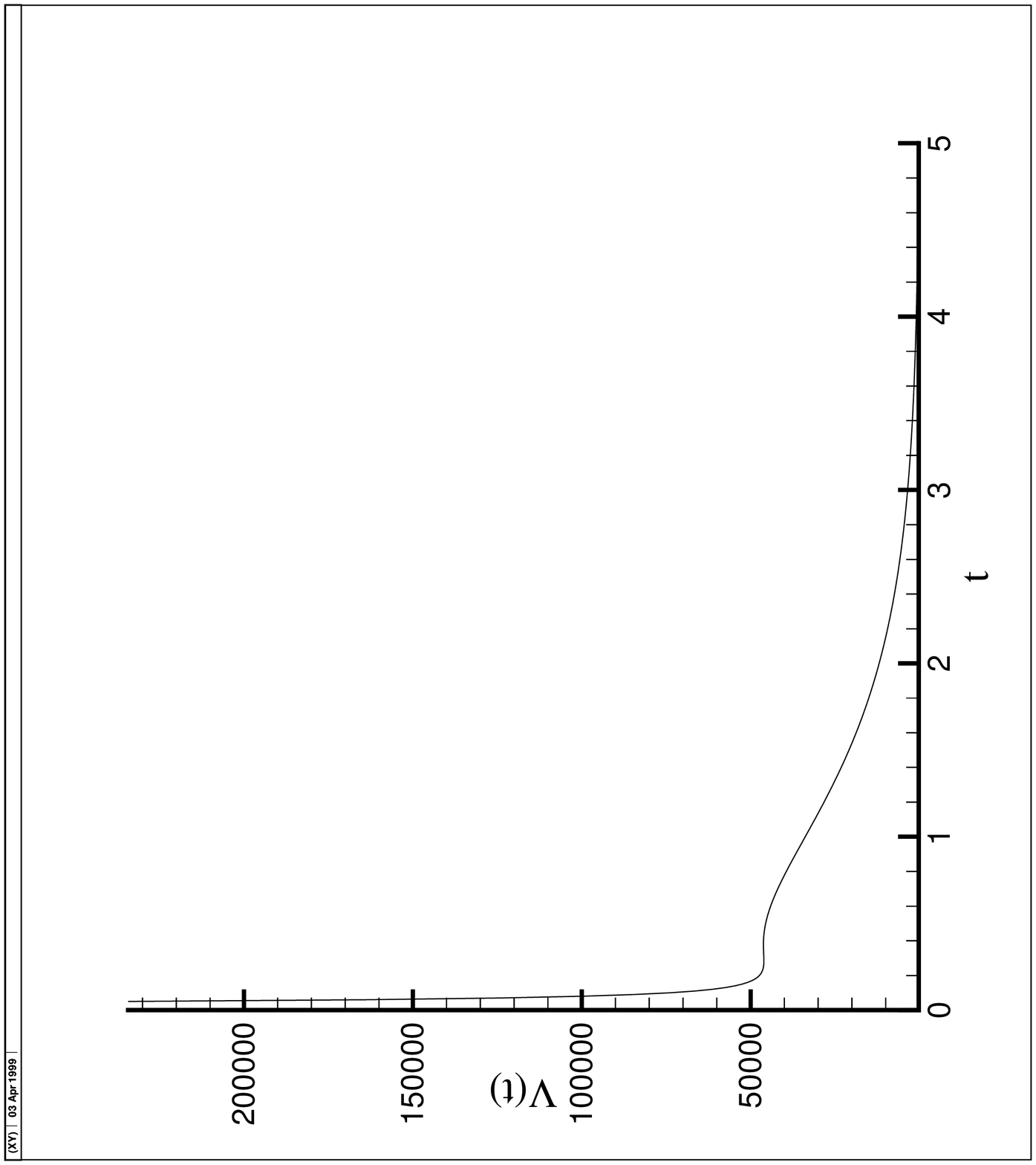,width=8cm}
\caption{ .}\label{fig4}
\end{figure}

The chois of parameters is special for a demonstration of the main effect
and equals to $\nu=6, ~u_0=30,~\mu=0.3, V_0=0, \gvk=1$.
The change of $\phi=\phi(t)$ and $V=V(t)$ versus time are presented in the
figures 5 and 6 using the parameters above.
\begin{figure}
\centering
\psfig{file=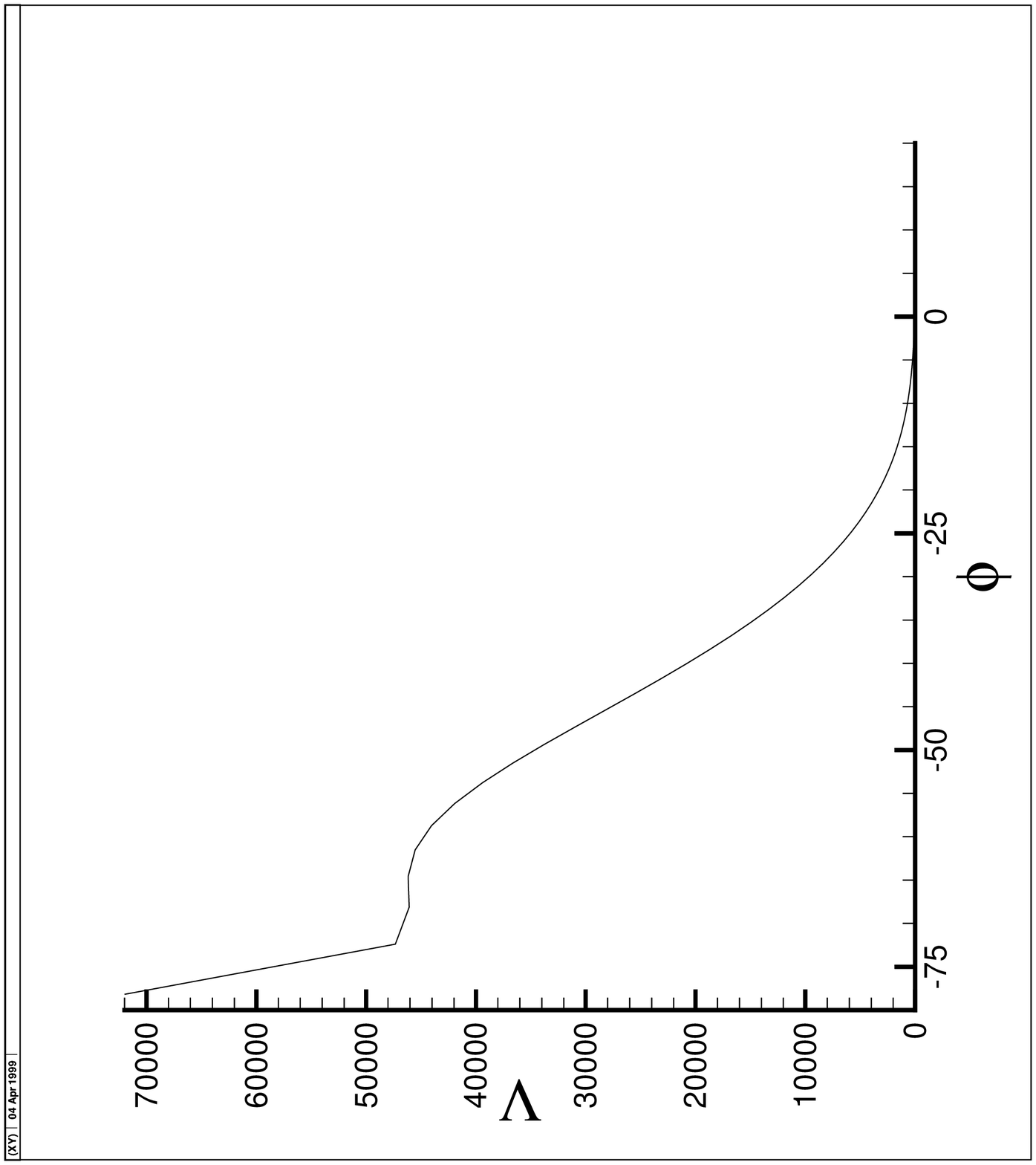,width=8cm}
\caption{ .}\label{fig5}
\end{figure}
\begin{figure}
\centering
\psfig{file=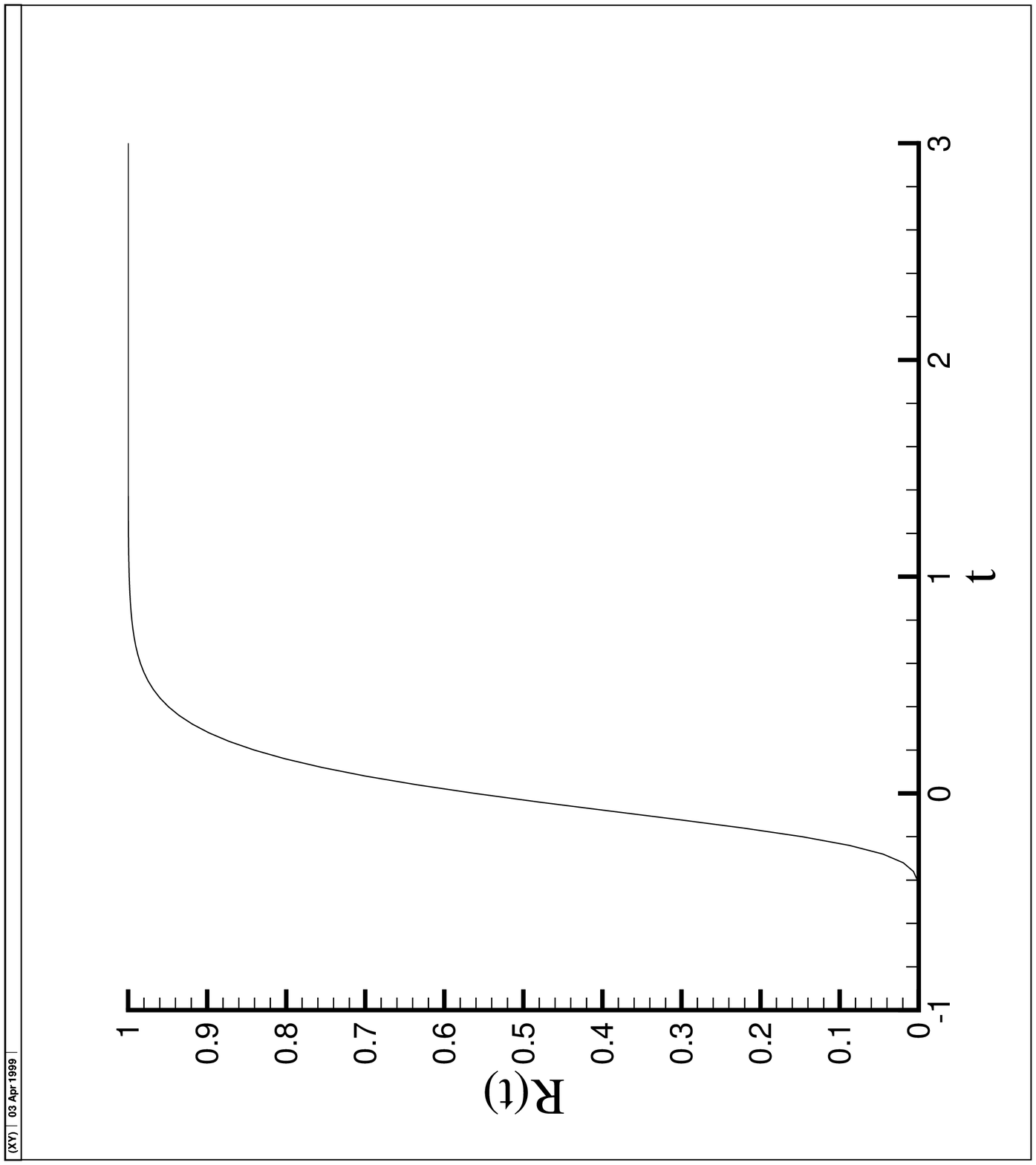,width=8cm}
\caption{ .}\label{fig6}
\end{figure}

The plot of dependence $V=V(\phi)$, calculated by plots of dependence
$\phi=\phi(t)$ and $V=V(t)$, is presented in the Fig. 7.
\begin{figure}
\centering
\psfig{file=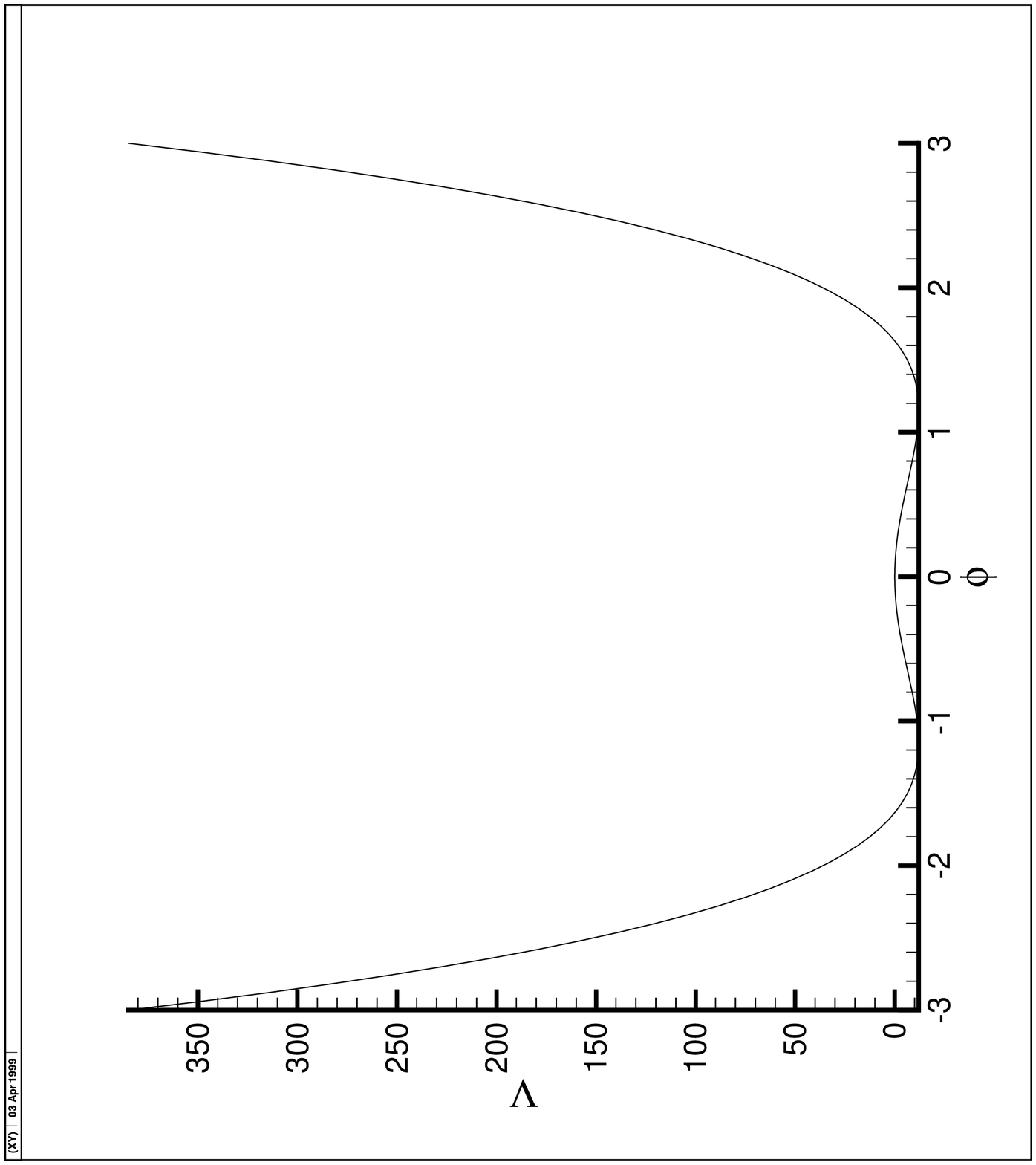,width=8cm}
\caption{ .}\label{fig7}
\end{figure}

Thus during the whole evolution the potential is a positive function
and conseiquently the energy dominant condition is true.

\section{Conclusion}\label{s5}

We have found new classes of exact solutions for a self-consistent system
of gravitating scalar fields with self-inter\-action within the framework
of the cosmology of a homogeneous isotropic universe.  We have analyzed
aspects of the given model that have bearing on the determination of the
physical conditions restricting the admissible form of the
self-inter\-action potential and on the determination of the general form
of evolution of the scale factor from the standpoint of the existence of an
inflationary phase in the evolution of the universe, along with other
fundamental characteristics of the potential.

1.  We have shown that the definition of the slow-roll regime actively
utilized in inflation is amenable to a variational formulation.  In this
formulation we have obtained an exact solution for the standard model of
inflation with a self-inter\-acting scalar field and have given an equation
of state of matter corresponding to this solution.

2.  We have proposed a method for generating exact solutions with a fixed
history of the potential $V = V(t)$.  On the basis of the method we have
formulated an eigen\-func\-tion/eigen\-value problem for a Friedmann flat
universe with the role of the eigenvalues taken by the cosmological
constant.  We have shown that in order to analyze the character of the
evolution of the universe, it is simpler to analyze the model dynamics with
the potential represented in the form of its history $V = V(t)$ than when
it is written as a function of the field $V = V(\phi )$.

3.  We have analyzed several characteristic types of potential history and
the corresponding evolutions of the scale factor.  Our analysis shows that
the formal analog of the Einstein equations in the form of the
Schr\"{o}dinger equation (\ref{e15}) or (\ref{e16}) as proposed in this
paper can be used to analyze in detail the behavior of various physical
factors in self-inter\-acting scalar field models and to discern a
potential selection criterion proceeding from physical notions regarding
the character of the evolution of the universe in large time scales,
including inflation (subinflation) as one of the stages, along with the
conditions for transition of the universe into a Friedmann regime.  It
follows from our analysis that the most realistic model of the history of
the potential energy is a model of the form~(\ref{e25}).

4.We have constructed the model of the evolution of homogeneous and
isotropic universe posessing an inflationary stage with an analitical exit
to the radiation dominated epoch and matter dominated era.
The model is constructed as a superposition of two auxiliary models,
the first of them is responsible for an inflationary stage, while the second
-- for the Freidmann's regim including matter dominated epoch.

This work has been carried out with partial financial support
from the Russian Foundation for Basic Research (Grant No. 98-02-18040).


\section*{References}

\begin{thebibliography}{99}
%
\bibitem{linde90} A. D. Linde, {\it Particle Physics and Inflationary Cosmology}
(Harwood Acad. Publ., Paris--New York, 1990) [Russ. original, Nauka,
Moscow, 1990].
\bibitem{guth81}  %guth81
A. H. Guth, Phys. Rev. D {\bf 23}, 347 (1981).
%
\bibitem{csz96iv}
S. V. Chervon, Shchigolev V.K. and V. M. Zhuravlev,
Izv. Vyssh. Uchebn. Zaved. Fiz., N 2, 1996, 41.
%
\bibitem{ch97m} S. V. Chervon, {\it Nonlinear Fields in the Theory of
Gravitation and Cosmology} [in Russian] (Izd. Srednevolzhsk. Nauchn.
Tsentra, Ulyanovsk, 1997).
\bibitem{czs97}
Chervon S.V., Zhuravlev V.M., Shchigolev V.K.
%{\it New exact solutions in standard inflationary models.}
Phys.Let. {\bf B 398}, 269 (1997).
%
\bibitem{ch97gc}
S.V.Chervon,
%{\it Gravitational Field of the Early Universe I: Non-linear scalar field
%as the source}//
Gravitation \& Cosmology, v.3, No.2, p.145-150, 1997.
%
\bibitem{ZCS98} V. M. Zhuravlev, S.~V. Chervon and V. K. Shchigolev,  JETP,
{\bf 114}, N 2, 179 (1998).
%
\bibitem{zelnov75}
Ya.B.Zeldovich, I.D.Novikov,
{\it Structure and evolution of the Universe}//
Мoscow.:Nauka.-1975.-736 с.
%
\bibitem{mtr95}
R. Maartens, D. R. Taylor, and N.~Roussos, Phys. Rev. D {\bf52}, 3358 (1995).
%
\bibitem{burbar88}
A.B.Burd, J.D.Barrow,
%{\it Inflationary models with exponential potentials.}//
Nucl.Phys., {\bf 308}, 929 (1988) .%-945.
%
\bibitem{gmss92}
S.Gottl\"ober, V.M\"uller, H.-J.Schmidt and A.A.Starobinsky,
Int.J.of Modern Phys. D, {\bf 1}, No.2, 257 (1992) .
%
%\bibitem{ХокЭл} С.Хокинг, Дж.Эллис, {\it Крупномасштабная структура
%пространства-времени}. М.: Мир (1977).
%
\bibitem{chezhu96iv} S. V. Chervon and V. M. Zhuravlev, Izv. Vyssh. Uchebn. Zaved.
Fiz., No.\,\,8, 81 (1996).
\bibitem{3} P. Coles and F. Lucchin, {\it Cosmology: the Origin and
Evolution of Cosmic Structure} (Wiley, Chichester, 1995).
\bibitem{5} R. H. Brandenberger, in {\it Physics of the Early Universe,
Proceedings of the 36th Scottish Universities Summer School in Physics},
edited by J.~A. Peacock, A.~F. Heavens, and A.~T. Davies (1989), p.~281.
\bibitem{6} V. N. Lukash and I. D. Novikov, in {\it Observational and
Physical Cosmology}, edited by F.~Sunchez, M.~Collados, and K.~Rebolo
(Cambridge Univ. Press, 1990).
\bibitem{7} S. V. Chervon and V. M. Zhuravlev, in {\it Abstracts of the
Reports at the International School-Sem\-inar ``Foundations of Gravitation
and Cosmology,'' Odessa} (RGS, Moscow, 1995), p.~67.
\bibitem{11} J. D. Barrow, Phys. Rev. D {\bf 49}, 3055 (1994).
\bibitem{mtr95} R. Maartens, D. R. Taylor, and N.~Roussos, Phys. Rev. D {\bf
52}, 3358 (1995).
\bibitem{13}P. Parsons and J. D. Barrow, Class. Quantum Gravity {\bf 12},
1715 (1995).
\bibitem{14} V. K. Shchigolev, V. M. Zhuravlev, and S.~V. Chervon, JETP
Lett. {\bf 64}, 71 (1996).
\bibitem{weinberg75} S. Weinberg, {\it Gravitation and Cosmology: Principles and
Applications of the General Theory of Relativity} (Wiley, New York, 1972)
[Russ. transl., Mir, Moscow, 1975].

\end{thebibliography}
\end{document}